# Iso-parametric tool path planning for point clouds

**Qiang Zou**[a, b] · **Jibin Zhao**[a, *]

[a] Shenyang Institute of Automation, Chinese Academy of Sciences, Liaoning, 110016, China;
[b] University of Chinese Academy of Sciences, Beijing, 100049, China.

**Abstract**   The computational consuming and non-robust reconstruction from point clouds to either meshes or spline surfaces motivates the direct tool path planning for point clouds. In this paper, a novel approach for planning iso-parametric tool path from a point cloud is presented. The planning depends on the parameterization of point clouds. Accordingly, a conformal map is employed to build the parameterization which leads to a significant simplification of computing tool path parameters and boundary conformed paths. Then, Tool path is generated through linear interpolation with the forward and side step computed against specified chord deviation and scallop height, respectively. Experimental results are given to illustrate effectiveness of the proposed methods.

## 1. Introduction

Free-form surfaces (e.g., aero-parts and molds) are widely used in manufacturing industries. And they are often machined by computer numerical control (CNC) machine tools that move its cutter or table along a specified trajectory. In the most common cases, the trajectory is so-called tool path which constitutes the core of computer-aided manufacturing (CAM). The automatic generation of tool path for free-form surfaces is a fundamental issue in modern CAD/CAM systems.

Planning tool path is a compromising between precision and efficiency, which mainly involves two aspects, path pattern and path parameters. The former is about which shape tool path is. More specifically, there are three path patterns so far: direction parallel, contour parallel and spiral. The latter concerns geometric parameters of tool path, i.e., forward step and side step bounding chord deviation and scallop height, respectively. In this paper, we shall use the terminology interval to refer to offsetting distance on surfaces and the step is used to refer to parametric offsetting distance. Fig. 1 shows the three path patterns and Fig. 2 describes the two path parameters. Tool path planning on a surface is closely related to its representations (e.g., spline surfaces, meshes and point clouds). The point cloud as a direct description of surfaces has been receiving a growing attention since the paper [1] by Marc Alexa et al. And a considerable part of meshes and spline surfaces encountered in CAD/CAM are reconstructed from point clouds through approximation. However, the reconstruction process is complicated and computational consuming. What's worse, it is non-robust especially when the point cloud is a non-uniform sampling. Therefore, the direct planning of tool path for point clouds is of great significance. Compared to mesh or spline surface based representations, the lack of topological information simplifies the representation and storage of surfaces. Yet, when it comes to the geometric processing (e.g., parameterization and differential properties), it becomes rather tough. That's why most algorithms of tool path planning for point clouds simply employ some junior geometric processing methods (e.g., intersection between parallel planes and the point cloud). It is hard to parameterization and estimating curvatures, as junior geometric processing, for point clouds without topological information, let alone planning tool path which belongs to senior geometric processing.

The basis for tool path planning was laid around 1990. For instance, iso-parametric method (Loney et al. [2]), iso-planar method (Huang Y. et al. [3]) and iso-scallop method (Suresh K. et al. [4], Lin RS et al. [5], Sarma R. et al. [6]) are some of the typical approaches. Some developments are, for example, iso-phote [7] and C-space [8] methods with tool orientation planning taken into account too. The former in fact proposed another method for parameterizing and the latter introduced the classical

C-space method for robotics into tool path planning. What should be noted is the work [9] by G. W. Vickers and K. W. Quan providing a mathematical method to determine the interval between consecutive paths (i.e., side step calculation). And [3] presented a true machining error calculation method with which accurate forward step can be determined. While there are enormous literatures focusing on tool path planning, these specified for point clouds are few. The earlier methods of direct planning on point clouds resorted to resampling, which can be regarded as an extension of traditional iso-planar ones. Lin et al. [10] employed a uniform rectangular grid (i.e., the Z-map constructed from an original point cloud) to generate tool path for milling surfaces slice-by-slice. Rows on each slice (or level) of the grid were picked out as tool path for the slice, with some segments trimmed to avoid islands. Another relevant method is the work [11] by Feng and Teng. They adaptively computed the forward and side step by the construction of a so-called CL-net which in fact is a variant of the Z-map. The iso-scallop method [6] for spline surfaces was extended to point clouds by Wei et al. [12]. Apart from those, S. C. Park et al. [13] generated contour parallel tool path for pocket milling from a data structure point of view (i.e., PSC-map). The preceding review is carried out regardless of the approaches based on reconstruction of meshes and spline surfaces from point clouds locally or globally which in fact steps backward. Surveys of much more work in tool path planning research can be referred in [14, 15]. Unfortunately, the extension of traditional iso-parametric tool path planning, one of the most important planning means, to point clouds seems to remain blank. From the above review of planning tool path on point clouds, it is evident that much more research is still needed to be carried out, especially for parametric methods that avoid the costly computations of surface-surface intersection and surface offsetting.

In this paper, a method of iso-parametric tool path planning for point clouds is presented. Surface parameterization, closely related to machine learning and computer graphics [16, 17], is crucial for planning tool path iso-parametrically. Yang et al. [18] and Sun et al. [19] employed the harmonic map to parameterize spline surfaces and meshes respectively, of which the free-boundary property was exploited to plan boundary conformed paths. The property means that boundary of parametric domain can be defined arbitrarily. Thus, by mapping spatial boundaries to regular planar boundaries (i.e., rectangles and circles), boundary conformed tool path can be generated. However their mesh-based mapping methods can't be applied to point clouds directly, as opposed to the employed method in this paper which is specified for point clouds by Schmidt and Singh [20]. Moreover, there is another exciting property that their work didn't cover, i.e., the angle preserving property. The advantage of conformal parameterization over conventional ones is the conformality (i.e., angle preserving) with which the computation of side and forward step can be simplified significantly. As known, a single path on a free-form surface is offset from a previous one along the direction orthogonal to forward (or feed) direction. As the conformal parameterization is angle preserving, the direction on a surface is consistent to that in its parametric domain. Therefore, for iso-parametric tool path, the forward direction is the $v$ -direction (or $u$ -direction) and the offsetting direction is the $u$ -direction (or $v$ -direction). What's more, the step size calculation is avoided if the parameterization is conformal, since for any point on the surface, its local shape is similar to its parametric image, with respect to a factor. Namely, a small interval on the surface and its corresponding step on the parametric domain are proportional. The assumption that each point of tool paths should be one of the existing points of a point cloud is unreasonable. Thus, interpolation is inevitable. The one we choose is of linear precision but much simpler than rational spline interpolation. It allows a point to be expressed as a weighted linear combination of its neighbor points. What's more, it is consistent to the conformal parameterization. In fact, these presented approaches can also be exploited to generate iso-scallop tool path for point clouds as an extension of the work [4, 5].

The remaining parts of this paper are:

Section 2 introduces the conformal parameterization for point clouds;

Section 3 presents the simplification of path parameters calculation;

Section 4 describes the proposed iso-parametric tool path planning;

Section 5 shows experimental results;

Section 6 concludes the paper.

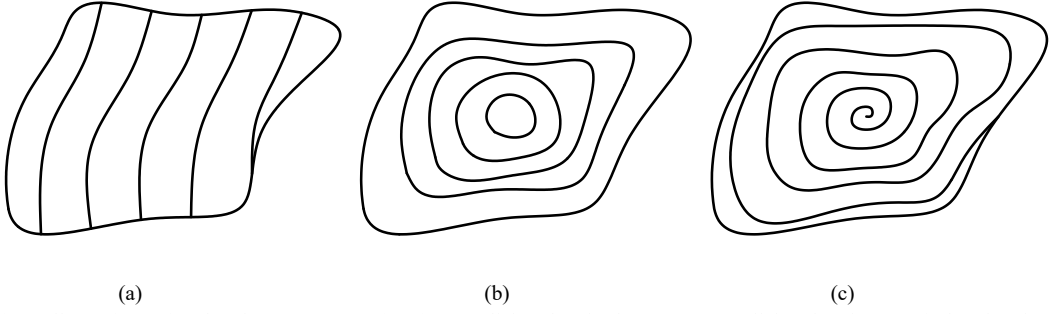

Fig. 1 Illustrations of tool path patterns. (a) Direction parallel tool path; (b) Contour parallel tool path; (c) Spiral tool path.

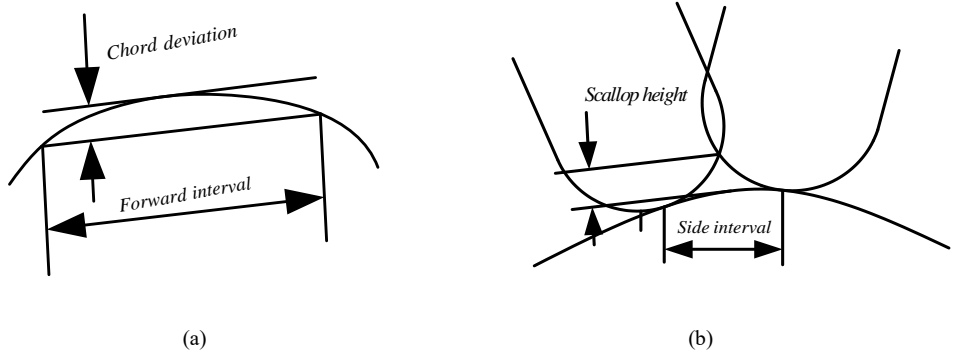

Fig. 2 Definition of tool path parameters.

## 2. Conformal point cloud parameterization

In continuous cases, parameterizing a surface $\Sigma$ embedded in $R^3$ can be modeled as

$$\varphi : D \to S, D \subseteq R^2,$$

where $D$ is a planar domain. An alternative method for conformal parameterizing via harmonic map instead of conformal map is much simpler. A harmonic map satisfies the Laplace equation

$$\nabla^2 f = 0 ,$$

with respect to the Dirichlet boundary condition: $f : \partial\Sigma_1 \to \partial\Sigma_2$ is homeomorphism, where $\nabla^2$ donates the Laplacian and $f$ is a map over one surface $\Sigma_1$ to the other $\Sigma_2$. If $\partial\Sigma_2$ is planar, the Laplace equation has a unique solution. Thus, its inverse is a conformal parameterization for the surface $\Sigma_1$. Similarly, in the cases of point clouds, such parameterization can be obtained by constructing discrete Laplacian and solving discrete Laplace equation.

### 2.1 Discrete Laplace equation

The discrete Laplacian $L$ for a point cloud $S = \{p_1, p_2, \cdots, p_n\} \subseteq R^3$ is a linear operator which takes a map on $S$ as input and another map on $S$ as output. If the size of $S$ is $n$, a map on $S$ is an $n$-dimensional vector. Therefore, $L$ is a linear map between two $n$-dimensional vector spaces, which can be represented as a $n \times n$ matrix. Generally, most methods for constructing such matrix fall into two categories, the Graph based methods and the Finite Element Method (FEM) based methods.

The graph Laplacian $L$ is defined as

$$l_{ij} = \begin{cases} \deg(p_i) & i = j \\ -1 & i \neq j \ and \ p_j \ neighbor \ to \ p_i \\ 0 & otherwise \end{cases} ,$$

where $\deg(p_i)$ donates the degree of point $p_i$. And subsequently, the Laplacian $L(f(p_i))$ of a map $f$ at a point $p_i$ can be expressed as a local sum

$$(Lf)(i) = \sum_{p_j \in N(p_i)} \left[ f(p_i) - f(p_j) \right],$$

where $N(p_i)$ is a set of points neighbor to $p_i$, e.g., the K nearest neighborhood (KNN). As for the FEM based methods, it is defined as

$$(Lf)(i) = \sum_{p_j \in N(p_i)} w_{ij} \left[ f(p_i) - f(p_j) \right],$$

where $w_{ij}$ donate the introduced weights. In the literature of parameterization, there have been several weights presented for meshes. For example, the famous cotangent scheme proposed by Pinkall et al. [21] and the mean-value scheme proposed by Floater [22]. Since most of the weights were derived on triangular meshes, it is hard for them to be applied to point clouds directly. However, Floater and Reimers [23] were able to extend their weights to a point cloud, which needs local triangulating. A weight scheme specified for point clouds has recently been proved to converge to the continuous Laplacian as sampling getting denser, Belkin and Niyogi [24]. The weights are as follows

$$w_{ij} = e^{-\frac{\|p_i - p_j\|}{t}}, t \in R^+ \text{ and } p_j \in N(p_i),$$

where parameter $t$ is constant for a point cloud. However, it is very hard to choose a proper $t$ for a real model. There is another weight scheme (optimal weights) coinciding with Belkin's [24]. Schmidt and Singh [20] parameterized point clouds conformally with such optimal weights which minimizes a quadric error

$$\varepsilon = \left\| p_i - \sum w_{ij} p_j \right\|^2, \ \ p_j \in N(p_i),$$

subjecting to a constraint $\sum w_{ij} = 1$. The error $\varepsilon$ can be rewritten as

$$\varepsilon = \left\| \sum w_{ij}(p_i - p_j) \right\|^2.$$

Then, minimizing the error becomes a least-square problem. Re-donate the indices of neighbor points as $\{1, \cdots, m\}$, where $m$ is the number of neighbor points. The solution is to solve a linear system

$$CW = 1, \ \ c_{jk} = (p_i - p_j)^T (p_i - p_k),$$

where 1 is the one-vector. And rescale the weights so that they concide with the constraint.

Once the discrete Laplacian for a point cloud has been constructed, the Laplace equation for a point cloud becomes $Lf = 0$. Suppose that $\partial S = \{p_{r+1}, \cdots, p_n\}$, $\partial D = \{q_{r+1}, \cdots, q_n\}$ and the boundary map $f : p_i \mapsto q_i$, $r < i \le n$. Then the conformal parameterization problem comes to solve a linear system

$$Lf = 0 \Rightarrow AU^I = -BU^B, \tag{1}$$

where

$$\begin{bmatrix} A_{n \times r} & B_{n \times (n-r)} \end{bmatrix} = L, \ U^I_{r \times 2} = [q_1 \cdots q_r]^T \text{ and } U^B_{(n-r) \times 2} = [q_{r+1} \cdots q_n]^T.$$

However, in continuous cases, there is no differential property for boundary points. Therefore, rows associated with boundary points should be removed. Namely,

$$\begin{bmatrix} A_{r \times r} & B_{r \times (n-r)} \\ C_{(n-r) \times n} \end{bmatrix} = L.$$

This can also be rewritten as

$$q_i - \sum_{p_j \in N(p_i) \backslash \partial S} w_{ij} q_j = \sum_{p_j \in N(p_i) \cap \partial S} w_{ij} q_j, \ \ i = 1, 2, \cdots, r.$$

Note that $q_i = [u_i, v_i]^T$ and the equation (1) is solved twice, one for $u$ and the other for $v$ coordinate.

## 2.2 Boundary points mapping

The pre-step for constructing conformal parameterization is to define boundary map, with mainly two aspects involved, shape of planar boundary and distribution of boundary points. When planning contour parallel tool path, the boundary is mapped to a circle in a Polar coordinates system. Thus concentric circles (i.e., iso-curves with respect to radius) are corresponding paths in parametric domain, which are free from the intersection detection and trimming, as opposed to conventional offsetting methods. Paths are generated in a manner as they morph inward from the boundary and get smoother along the morphing. When planning direction parallel tool path, the boundary is mapped to a rectangle in a Cartesian coordinates system, so that the paths can be simply generated by selecting segments parallel to either of the two edge pairs. What's more, this regular domain can help to avoid paths of small size (i.e., not boundary conformed), if the underlying surface is trimmed, as opposed to conventional iso-planar methods. This merit was shown by Yang et al. [18]. As for distribution, there are two procedures involved, ordering points and assigning them.

The method used to order boundary points is inspired by the material in Floater's [22]. First, a point cloud is classified into two subsets, interior and boundary. Then, these boundary points are manually broken into several simple parts, as shown in Fig. 3(a). For each part, boundary points are parameterized into the unit interval [0,1] by the method similar to section 2.1. Their boundary points are the end points of each part (i.e., the breaking points). And their weights are

$$w_{ij} = 1 / \| p_j - p_i \|, \ \ p_j \in N(p_i) .$$

The ordering of the parameter values is used to order the 3D boundary points. Finally, all parts are combined into an ordered boundary according to the ordering of breaking points.

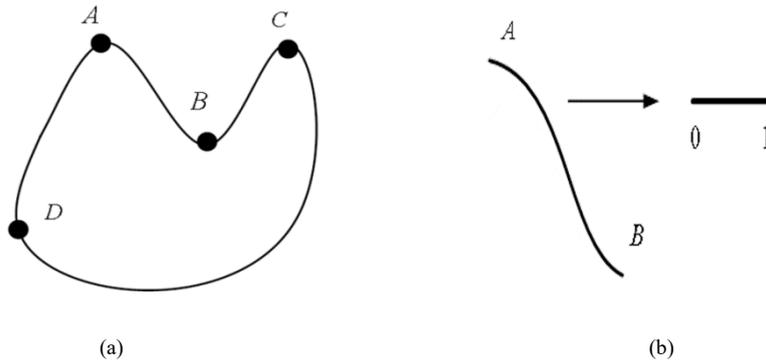

(a)                                                                 (b)

Fig. 3 Ordering boundary points. (a) Boundary breaking; (b) Bounadry parameterization.

The boundary points are assigned along the planar boundary according to chord length between adjacent points on the 3D boundary. Specifically, when planning contour parallel tool path, boundary points are mapped to a circle in a Polar coordinates system with

$$f : p_i \mapsto (R_0, \lambda_i \cdot 2\pi), \ \ p_i \in \partial S ,$$

Where radius $R_0$ can be set arbitrarily and we set it to be $L/(2\pi)$ with $L$ being length of the ordered boundary. $\lambda_i$ is the distribution parameter with $\lambda_{r+1} = 0$ and for $r+1 < i \le n$

$$\lambda_i = \sum_{j=r+2}^{i} \| p_j - p_{j-1} \| \Big/ L .$$

Intuitively, points are proportionally mapped around the whole boundary, with respect to chord length.

When planning direction parallel tool path, the boundary points are mapped to a rectangle. They are first divided into four parts manually, so that four breaking points are consistent to the four planar corner points. Inherently, each part has a local ordering. Then, each part is mapped to its corresponding planar edge proportionally. Suppose that consecutive lengths of the four parts are $L_1$, $L_2$, $L_3$ and $L_4$. Then the lower left point is set to be $(0,0)$ and the upper right point is $\big( (L_1 + L_3)/2, \ (L_2 + L_4)/2 \big)$.

The first part is mapped to the $u$ -axis with

$$f : p_i \mapsto (\lambda_i \cdot (L_1 + L_3) / 2, 0), \quad p_i \in first \ part \ of \ \partial S \ ,$$

where $\lambda_i$ is the assigning parameter with $\lambda_{r+1} = 0$ and for the rest points

$$\lambda_i = \sum_{j=r+2}^{i} \left\| p_j - p_{j-1} \right\| \Big/ L_1 \ .$$

And correspondingly for the rest three parts.

As the preceding shows, conformal point cloud parameterization can be boiled down to solve a sparse linear system. And it provides some perfect properties for tool path planning. The angle preserving property helps to simplify the computation of tool path parameters, the free-boundary property provides natural parametric domains for planning tool path iso-parametrically and the shape preserving property makes the 3D shape of paths consistent to that of 2D parametric ones.

## 3. Path parameters calculation

Determining the geometric parameters for tool path (i.e., forward and side step) is closely related to differential properties of a surface. Some basic notions about differential geometry are introduced and it is also shown how to simplify the computing of forward and side step in this section.

### 3.1 Differential geometry

There are many curvatures for a surface (e.g., Gaussian curvature, mean curvature and principle curvatures). And the one involved with tool path planning is normal curvature. Normal curvature is defined as the curvature of the curve that is the intersection between the surface itself and a plane containing both the normal vector $n$ and a direction vector $e$ on the tangent plane. Consider a parametric surface $\Sigma = \{ r \in R^3 | r = r(u,v), (u,v) \in D \subseteq R^2) \}$, the normal curvature is expressed as

$$\kappa = \frac{\mathrm{II}}{\mathrm{I}} = \frac{Ldu^2 + 2Mdudv + Ndv^2}{Edu^2 + 2Fdudv + Gdv^2} \ ,$$

where $E = r_u \cdot r_u$, $F = r_u \cdot r_v$, $G = r_v \cdot r_v$ are the coefficients of the first fundamental form $\mathrm{I}$ and $L = n \cdot r_{uu}$, $M = n \cdot r_{uv}$, $N = n \cdot r_{vv}$ are the coefficients of the second fundamental form $\mathrm{II}$. The pair $(du, dv)$ represents the direction $e$. For instance, the normal curvature along the direction $(du, 0)$ is $\kappa = L / E = n \cdot r_{uu} / r_u \cdot r_u$.

If the parameterization is conformal, it can be derived that $E = G$ and $F = 0$ which implies that directions $(0, dv)$ and $(du, 0)$ are orthogonal on the surface. What's more, the local shapes around $p \in \Sigma$ and $q \in D$, where $p = r(q)$, are similar with respect to a factor $\sigma$. The factor can be derived with

$$\sigma = \left| \frac{dr}{dv} \right| = \frac{\sqrt{\mathrm{I}}}{|dv|} = \frac{\sqrt{r_v \cdot r_v dv^2}}{|dv|} = |r_v| = |r_u| \ .$$

Therefore, a small increment on a surface is $|r_u|$ times of the corresponding parametric increment.

For discrete cases, the difficulty is an approximation of the first and second order differential properties and the unit normal estimation. The first idea may be rational spline surface reconstruction and analytical evaluation. However, this is rather complicated. We next use a well-known difference scheme to approximate the two properties. Consider a planar point $q_0$ and its KNN. Their corresponding spatial points can be obtained with the previous parameterization. Insert two points $q_1$ and $q_2$ into the KNN, as in Fig. 4. Their corresponding spatial points are computed using the method described in section 3.3 and the $\Delta u$ is chosen as the shortest distance between $q_0$ and its KNN. Expand $p_1$ and $p_2$ as the Taylor series

$$p_1 = r(u_0 - \Delta u, v_0) = r(u_0, v_0) - r_u(u_0, v_0) \cdot \Delta u + r_{uu}(u_0, v_0) \cdot \Delta u^2 + \mathrm{O}(\Delta u_1^{\ 3}) \ ;$$

$$p_2 = r(u_0 + \Delta u, v_0) = r(u_0, v_0) + r_u(u_0, v_0) \cdot \Delta u + r_{uu}(u_0, v_0) \cdot \Delta u^2 + \mathrm{O}(\Delta u_2^{\ 3}) \ .$$

Then

$$r_u'(u_0,v_0) = \frac{p_2 - p_1}{2\Delta u} \quad \text{and} \quad r_{uu}'(u_0,v_0) = \frac{p_2 + p_1 - 2p_0}{2\Delta u^2} \,.$$

And correspondingly for $r_v$ and $r_{vv}$. Note that the coefficients $F$, $M$ are not needed in our method.

As to the unit normal estimation, it can be regard as a local tangent plane approximation problem which in turn becomes a least-square fitting one. Then, it can be reduced to a spectrum analysis of the local covariance matrix [1]. Specifically, for a point $p_i$ and its KNN, the local covariance matrix is

$$C = \frac{1}{|N(p_i)|} \sum_{p_j \in N(p_i)} (p_j - c_i)(p_i - c_i)^T$$

where $c_i$ is the barycentric point. The optimal unit normal is the normalized eigenvector corresponding to the smallest eigenvalue.

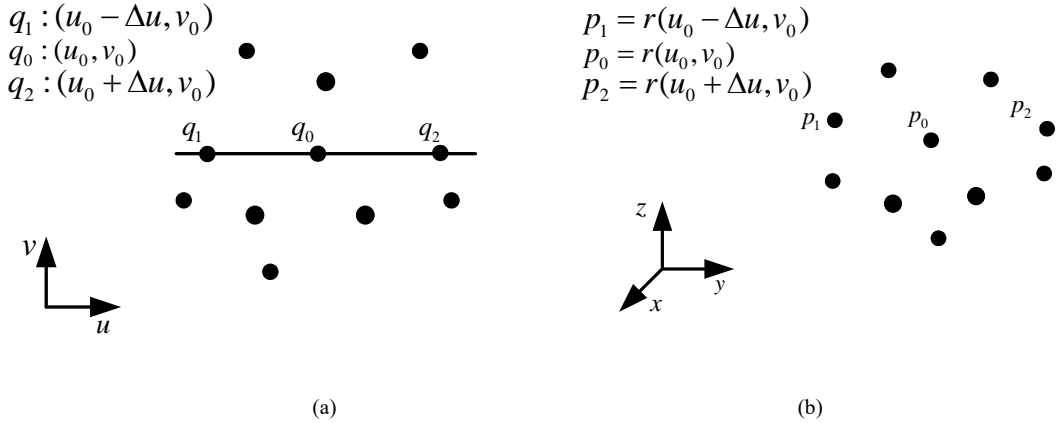

(a)                                                          (b)

Fig. 4 Differential properties approximation. (a) Planar points; (b) Spatial points.

## 3.2 Forward and side step

Forward step is responsible for chord deviation. A single tool path is usually discretized to be a series of line and arc segments (this paper focuses on line segments) for the limited capacity of CNC interpolator. The chord deviation is defined to describe the error of approximating a curve segment with a line one, details see [3]. Maximizing the forward step which determines the length of each segment is significant for machining efficiency.

The forward surface interval for machining can be expressed as

$$l_f = \sqrt{8eR - 4e^2} \,,$$

where $R$ is the normal curvature radius along the forward direction of a point and $e$ is the chord deviation. For iso-parametric tool path, the forward normal curvature radius along the direction $(0, dv)$ can be simplified as

$$R = \left| \frac{1}{\kappa_f} \right| = \left| \frac{\mathrm{I}}{\mathrm{II}} \right| = \left| \frac{G}{N} \right| = \left| \frac{r_v \cdot r_v}{n \cdot r_{vv}} \right| \,.$$

To generate a single iso-parametric tool path, the forward step should be computed. As mentioned, conformal maps preserve shapes infinitesimally, namely, the 3D shape is similar to the 2D counterpart in a small range. Therefore, locally, the forward step is proportional to the forward surface interval with respect to a factor $\sigma$ implying

$$\Delta v_i = v_{i+1} - v_i = \frac{l_{fi}}{\sigma_i} = \frac{l_{fi}}{|r_v|} \,.$$

Eventually, the forward step can be iteratively computed with the expression

$$v_{i+1} = v_i + \frac{l_{fi}}{|r_v|}. \tag{2}$$

Side step is responsible for scallop height. For machining free-form surfaces, they are often discretized to be a series of paths. The interval between two consecutive paths determines the scallop height that represents the error of approximating a surface with curves. Maximizing the side step is significant for machining efficiency too.

The side surface interval for machining can be expressed as

$$l_s = \sqrt{8hr\frac{R}{R+r}} \quad \text{or} \quad l_s = \sqrt{8hr\frac{R}{R-r}},$$

where $R$ is the normal curvature radius along the side direction of a point, $h$ is the scallop height and $r$ donates the ball-end cutter radius (so that tool orientation doesn't matter). The former expression is for convex surfaces and the latter is for concave ones. The side direction is orthogonal to forward direction on the surface, which is consistent to the parametric direction $(du,0)$, i.e., $\kappa_s = \kappa_u$ if the parameterization is conformal. Similar to forward step, the side normal curvature radius along the direction $(du,0)$ can be simplified as

$$R = \left|\frac{1}{\kappa_s}\right| = \left|\frac{\mathrm{I}}{\mathrm{II}}\right| = \left|\frac{E}{L}\right| = \left|\frac{r_u \cdot r_u}{n \cdot r_{uu}}\right|.$$

To generate the next iso-parametric tool path, the side parametric interval (i.e., side step) should be computed. Locally, for conformal parameterization and iso-parametric tool path, the side parametric interval is

$$\Delta u_i = u_{i+1} - u_i = \frac{l_{si}}{\sigma_i} = \frac{l_{si}}{|r_u|}.$$

Eventually, the side step for each point can be iteratively computed by

$$u_{i+1} = u_i + \frac{l_{si}}{|r_u|}. \tag{3}$$

Note that, in the cases of point clouds, the $|r_u|$ and $|r_v|$ are only approximations of the continuous cases where $\sigma = |r_u| = |r_v|$. Therefore, the mean value of them is taken as the similarity factor for point clouds.

### 3.3 Non-conformality error analysis

According to the Riemann Mapping Theorem, if the boundary shape of parametric domain is given, a conformal parameterization always exists in continuous cases. However, in discrete cases, mapping a complex 3D boundary to either a rectangle or a circle will inevitably cause distortion of conformality at points near boundary, since the Laplacian and boundary map are approximations. But the distortion is limited to points near boundary, as shown in Fig. 9 (a) (b). There are some ways to avoid this by computing boundary map as part of the solution, with a trade-off being that the free-boundary property is lost. However, as demonstrated in [18, 19], such property is of great significance for planning boundary conformed tool path. We next show the effect of such distortion on tool path.

For a point on the surface, its local shape is linearly related to its image in the parametric domain, with respect to a Jacobian. Suppose that side interval vector $\bar{l}_s = k_1 r_u + k_2 r_v$. Its corresponding parametric increment is

$$[\Delta u \ \Delta v] = J^{-1}(k_1 r_u + k_2 r_v) = [k_1/|r_u| \ k_2/|r_v|],$$

where $J = (r_u \ r_v)_{3 \times 2}$ is the Jacobian at the point, and the coefficients are $k_1 = l_s/\cos(|\theta - \pi/2|)$, $k_2 = l_s/\sin(|\theta - \pi/2|)$. The angle $\theta = \text{angle}(r_u, r_v)$. Subsequently, take $u$ direction as side direction,

the side step can be computed with

$$\Delta u = \frac{\sqrt{8hr\,R_s/(R_s \pm r)}}{\cos\left(\left|\theta - \pi/2\right|\right) \cdot |r_u|},$$

where $R_s$ is the normal curvature radius along the direction orthogonal to the forward direction. According to the Euler Theorem, the side normal curvature is $\kappa_s = 2H - \kappa_f$, where $H$ donates the mean curvature. For a point cloud, it can be easily estimated by the method [25]. This is the original expression for computing side interval. If we still employ the expression (3) at non-conformal points, it gives

$$\Delta u' = \frac{\sqrt{8hr\,R_u/(R_u \pm r)}}{|r_u|}.$$

If $\Delta u' \leq \Delta u$, the simplified expression will give conservative paths, i.e., the tool path is less efficient but the precision is guaranteed. Otherwise, the simplified formula will give incorrect steps at the point. Although the Euler theorem can relate $R_u$ with $R_s$, The sign of $\Delta u' - \Delta u$ is undermined. However, if the normal curvature radius at the point is much greater than the cutter radius (i.e., $R/(R+r) \approx 1$) or the local surface is isotropic (i.e., the normal curvature is constant at the point), the sign is negative, namely, the expression (3) tends to generate denser tool path for points near boundary.

The difference between expressions (3) and the original one is a projection procedure

$$l_s^u = k_1 = l_s \big/ \cos\left(\left|\theta - \pi/2\right|\right).$$

Then, for iso-parametric tool path, the side step is

$$u_{i+1} = u_i + \frac{l_{si}^u}{|r_u|}. \tag{4}$$

Never can all point clouds satisfy the previous conditions, i.e., for a few point clouds, the effect of boundary mapping on tool path planning is uncertain. And actually, distortion caused by boundary mapping only appears at the points which are very close to boundary or boundary corners. Thus, in this paper, the expression (4) instead of (3) is employed for paths near boundary. According to the experiments we conducted, the number of these paths is 3-5. This is rather conservative. A better way maybe choose the expression according to a criterion $|\theta - \pi/2| < \eta$. However, we do not yet know of a principled way to choose the threshold for each point cloud.

### 3.4 Linear interpolation

Given a point and the forward direction, its next parametric forward point and side point can be obtained with expressions (2) (3) (4). To compute the corresponding 3D coordinates, interpolation is needed. In order to be consistent with the conformal parameterization, the one used in this paper is linear interpolation which can help to obtain linear precision [26].

Consider a planar point $q_0$ with its KNN being $\{q_1 \cdots q_n\}$, its corresponding spatial point $p_0$ is

$$p_0 = \sum_{i=1}^{n} w_i p_i, \quad \sum_{i=1}^{n} w_i = 1,$$

where $\{p_i\}$ are the corresponding spatial points of the planar KNN. Although there are many schemes of the weights, they are chosen as the optimal one as in the conformal parameterization.

## 4. Tool path planning

Planning tool path is to represent a surface with a series of curves against some error criteria (i.e., chord deviation and scallop height). Iso-parametric tool path consisting of $m$ parametric curves (e.g., $\{r(u_0, v), \cdots, r(u_m, v)\}$) are generated by keeping one parameter (e.g., $u$) constant. We next show how to generate such curves on a surface for the two parallel tool path patterns respectively.

### 4.1 Direction parallel tool path

In order to construct row-like curves on surfaces by the iso-parametric method, the parameters $u$ and $v$ should be variables of a Cartesian coordinates system. As mentioned, the conformal parameterization has a free-boundary property. Therefore, the parametric domain for planning direction parallel tool path is set as a planar block, which means that the boundary is mapped as a rectangle. If the lower left point of the rectangle is set to be original point and its adjacent edges to be $u$-axis and $v$-axis respectively, forward direction can be chosen as the positive direction of $v$-axis and side direction as the positive direction of $u$-axis.

For each path, start from an initial point $(u_i, 0)$, iteratively determine the next forward point $(u_i, v_j) \rightarrow (u_i, v_{j+1})$ by the expression (2) and linear interpolation until $v$-coordinate is out of range, $v_{j+1} > b$, and set the last $v_j$ to be $b$. The first initial point is $(0,0)$, all other initial points are determined by side step.

For consecutive paths, compute side step for each point on the previous path with expressions (3) or (4), where the expression (3) is for interior points and (4) for points near boundary. Then select the minimal step as side step for next path. It is a two-stage procedure. First plan 3-5 paths from each side with the expression (4) resulting in a narrowed rectangular parametric domain, which is shown by dotted segments in Fig. 5 (a). Then iteratively determine side step from $u_0$ to $u_1$, as shown by the solid segments in Fig. 5 (a).

Note that the tool paths are generated without offsetting boundary inward a cutter radius distance.

### 4.2 Contour parallel tool path

The difference between contour and direction parallel tool path is the coordinates system used. For contour parallel tool path, the parameters $u$ and $v$ should be variables of a Polar coordinates system. And the boundary is mapped as a circle with its center being the original point and $R$ the radius. Forward direction can be chosen as the positive direction of $\theta$-axis and side direction as as the positive direction of $\rho$-axis.

For each path, start from an initial point $(\rho_i, 0)$, iteratively determine next forward point $(\rho_i, \theta_j) \rightarrow (\rho_i, \theta_{j+1})$ by the forward step computing and linear interpolation until $\theta$-coordinate is out of range $2\pi$. And set the last $\theta_j$ to be $2\pi$.

For consecutive paths, similar to direction parallel tool path, it is divided into two parts. First plan 3-5 paths from the boundary, which also results in a narrowed circular parametric domain. Then compute side step for each path from the original point to the circle $R'$.

Note that the first path is a point (the original point). Its side step is select as the minimal one from the side steps calculated along different direction around it.

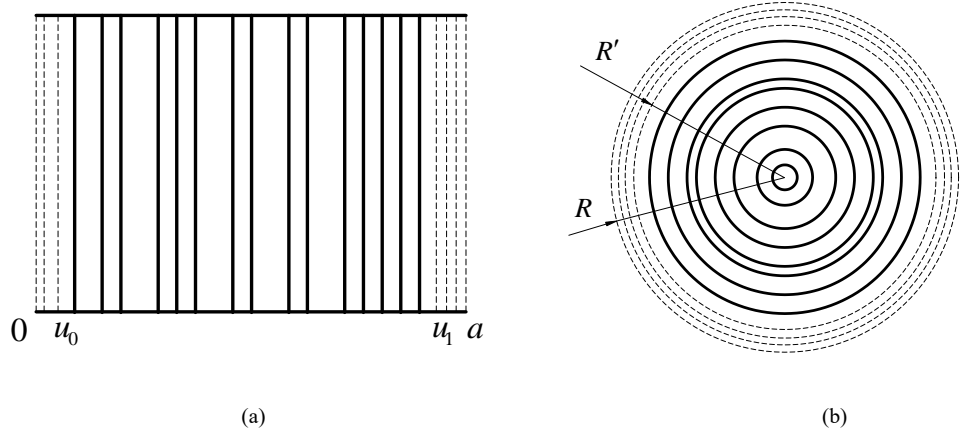

(a)                                          (b)

Fig. 5 Iso-parametric tool path planning. (a) Direction parallel tool path; (b) Contour paralle tool path.

## 5. Experimental results

In this section, the proposed method is implemented on real data. Three typical models are chosen to illustrate the effectiveness of it, as in Fig. 6. A cockpit surface (a) is used to plan direction parallel tool path, a human face (b) is used to generate contour parallel tool path and a free-form surface (c) with complex boundary is used to show both paths. The former two point clouds were generated by a 3D scanner (a coordinate measuring machine). The latter was generated by the UG software. Their ranges are $221.49 \times 105.83 \times 58.25$, $139.57 \times 118.93 \times 32.76$ and $83.50 \times 48.21 \times 18.46$, respectively.

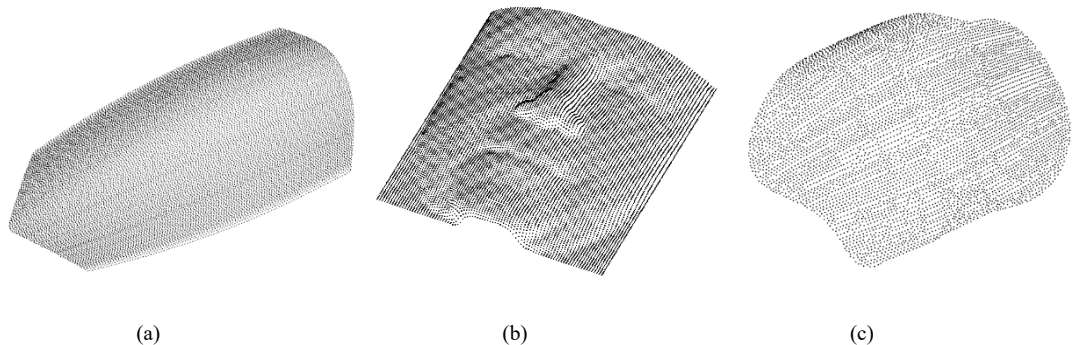

(a)         (b)         (c)

Fig. 6 Tested point cloud models. (a) The surface of a cockpit; (b) The surface of a human face; (c) A free-form surface.

For planning iso-parametric tool path, the first thing to do is parameterization. The K-d tree is exploited to quickly search for KNN and the number of neighbor points is 12. When ordering boundary points the number is 4 . The algorithm in [24] is adopted to extract boundaries from point clouds. The algorithm GMRES is chosen to solve the sparse equations. Note that the former two point clouds are smoothen by the Laplacian fairing method and the Laplacian scheme is consistent to those in section 2.1. The results of parameterization are presented in Fig. 7, Fig. 8 and Fig. 9. Fig. 7 (a) shows the angle preserving property of the cockpit point cloud with a rectangular domain. Fig. 8 (a) shows the angle preserving property of the face point cloud with a circular domain. Fig. 9 (a) and (b) shows the angle preserving property of the free-form point cloud with both domains.

As point clouds have become parametric ones, a series of forward and side points can be computed using the analytical expressions in section 3.2. And subsequently, the corresponding tool path can be generated by linear interpolation. A ball-end cutter with radius $r = 4mm$ is chosen to illustrate the path generation so that tool orientation doesn't matter. The limited scallop height is $h = 1mm$ and chord deviation is $e = 0.01mm$ . In order to clearly showing tool paths, the error criterion (scallop height) is set to be much greater than real cases. Fig. 7 shows the direction parallel paths on the cockpit point cloud. A comparison between proposed method and the conventional iso-planar method is also given. Fig. 7 (a) and (b) show paths generated by the iso-planar method. Fig. 7 (d) and (e) show corresponding paths by the proposed method. Fig. 8 shows the contour parallel paths on the face point cloud and a comparison between proposed method and the conventional offsetting method. Fig. 8 (b) shows paths generated by the proposed method. Fig. 8 (c) show corresponding paths by the offsetting method. Fig. 9 shows both path patterns planned for the free-form point cloud. Fig. 9 (c) shows direction parallel tool path and Fig. 9 (d) shows contour parallel tool path.

As the figures show, for direction parallel tool path, the lengths of paths are rather even and it is boundary conformed, as opposed to iso-planar methods which often generate uneven paths in terms of length if the initial plane is chosen poorly. And, for contour parallel tool path, boundary morphs inward gradually making tool path boundary conformed, as opposed to conventional offsetting methods needing the post-process of removing intersection between offsetting paths. Another problem of conventional contour parallel tool path is that they preserve sharp corners of boundaries, which limits the feed-rate when approaching the corners. And then, machining efficiency and tool wear are reduced. However, the method proposed rounds these corners automatically and gradually.

As mentioned, discrete boundary mapping will cause distortion near boundary. In Fig. 10, the effect of different assignments is shown, one assigns boundary according to chord length between adjacent 3D points (the adopted method) and the other assigns boundary uniformly. It is shown that chord length based method can give better results, especially when the 3D boundary points are irregularly distributed. In Fig. 11, the effect of corners of different angles on tool path is shown. These corners are mapped to be right angles when planning direction parallel tool path, which causes local non-conformality. The effect of these corners seems to make side intervals conservative when employing the simplified side step calculation formula, as shown in Fig. 11 (c). The model is chosen to illustrate the effect because its normal curvature radius is much greater than the cutter radius and it has three typical angles, obtuse, acute and right angles. The acute angle is $25.56°$ and the scallop height is limited to $h = 0.03mm$. The number of paths in (b) is 26 and 30 in (c). Fig. 12 shows side interval approximation error $\varepsilon(\%) = (|\, l_s - l_s' \,|/l_s) \times 100$ where $l_s' = \Delta u \cdot \sigma$ and $l_s$ donates side interval between two adjacent paths of the human face model (Fig. 8). Fig. 12 (b) shows error of two interior adjacent paths and (c) shows error of two paths near boundary. The mean value, maximum value and minimum value of (b) are 1.4083, 3.1971 and 0.0380 respectively. The corresponding values of (c) are 2.2442, 4.884 and 0.1945. This side interval error can be transferred to the error of scallop height $e$ with a simply linear expression, which is shown as follows

$$l_s = \sqrt{8hrR/(R+r)} \Rightarrow 2l_s\Delta l_s + \Delta l_s^2 = 8\Delta hrR/(R+r),$$

Divide both sides by $l_s^2$, having

$$2\Delta l_s / l_s = \Delta h / h \Rightarrow e = 2\varepsilon.$$

Therefore, the corresponding mean error of scallop height is around 5%.

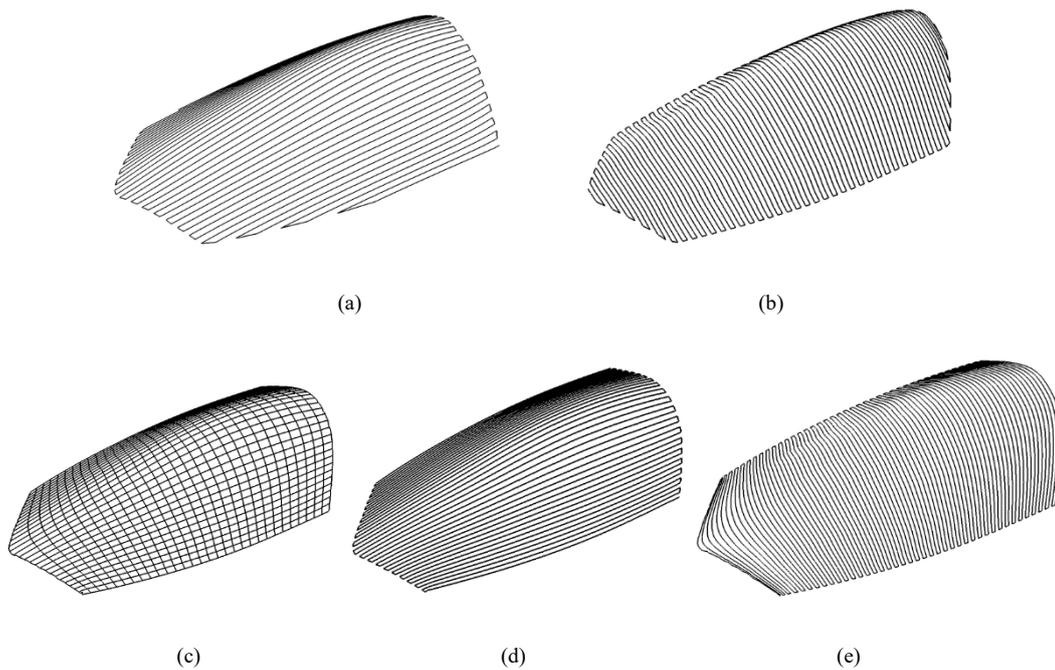

(a)                    (b)

(c)                    (d)                    (e)

Fig. 7 The results of a cockpit surface. (a) (b) Tool path generated by the iso-planar method; (c) Angle-preserving property with a rectangular domain; (d) Direction parallel tool path in one direction; (e) Direction parallel tool path in the other direction.

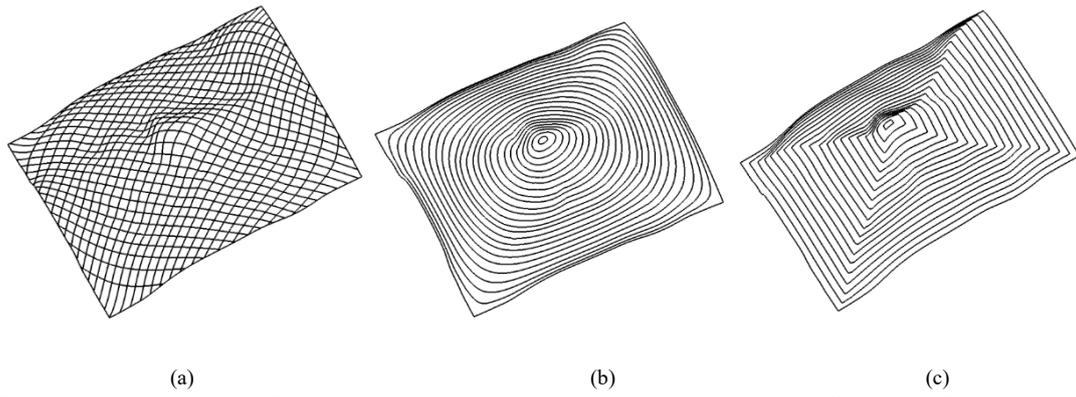

(a)                                    (b)                                    (c)

Fig. 8 The results of a face surface. (a) Angle-preserving property with a circular domain; (b) Contour parallel tool path; (c) Tool path generated by the conventional offsetting method.

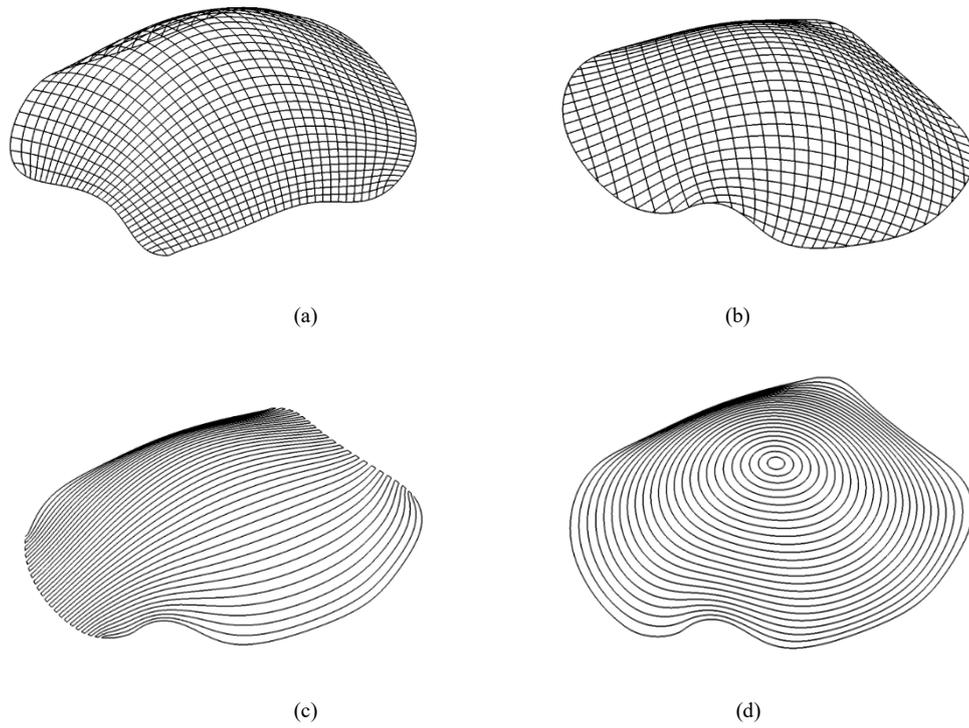

(a)                                    (b)

(c)                                    (d)

Fig. 9 The results of a freeform surface. (a) (b) Angle-preserving property with a rectangular domain and a circular domain, respectively; (c) Direction parallel tool path; (d) Contour parallel tool path.

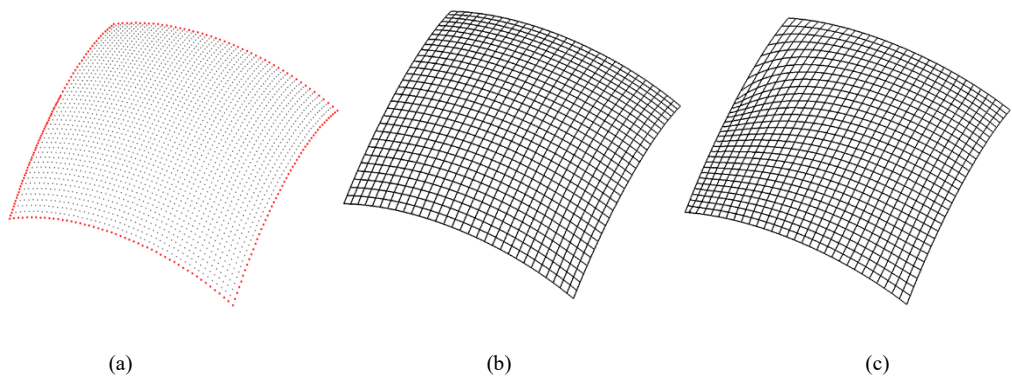

(a)                                    (b)                                    (c)

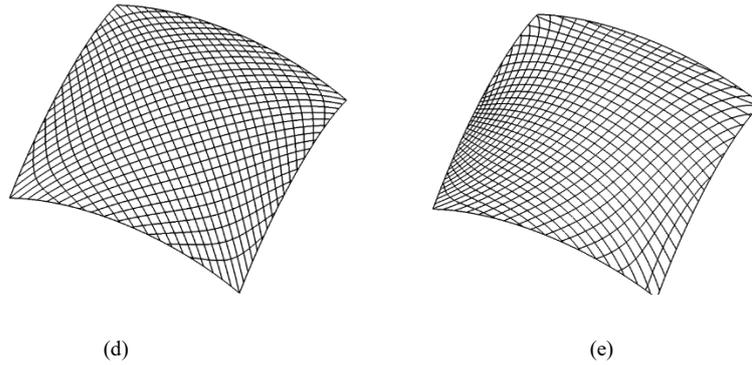

(d)                                      (e)

Fig. 10 Comparison of different assignments. (a) Original point cloud with irregular boundary; (b) Parameterization with a rectangular domain using chord length boundary assignment; (c) Parameterization with a rectangular domain using uniform boundary assignment; (d) Parameterization with a circular domain using chord length boundary assignment; (e) Parameterization with a circular domain using uniform boundary assignment.

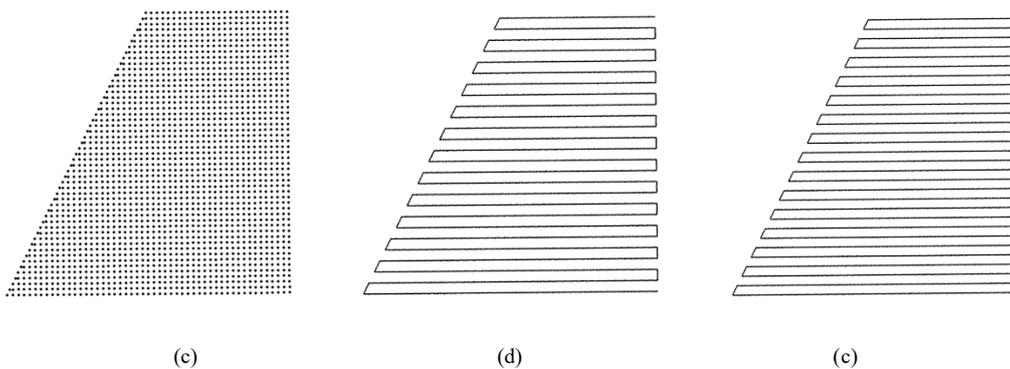

(c)                          (d)                          (c)

Fig.11 The effect of distortion near boundary on tool path. (a) Original point cloud with different angle; (b) Paths planned using the accurate expression (4); (c) Paths planned using the simplified expression (3).

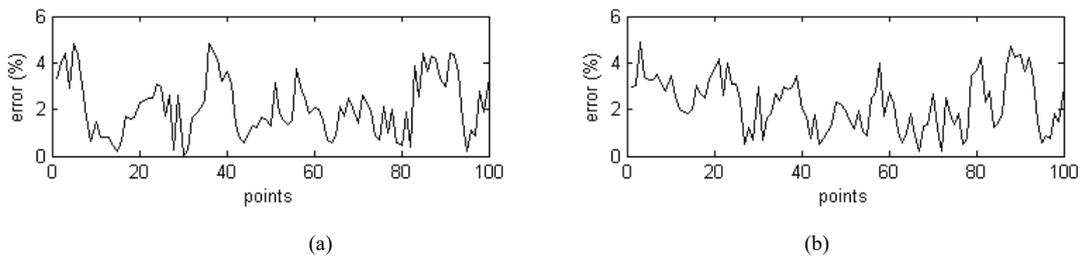

(a)                                      (b)

Fig. 12 The error caused by discrete conformal parameterization. (a) Side interval error between two adjacent interior paths of the face model; (b) Side interval error between two adjacent paths near boundary of the face model;.

## 6. Conclusions

The shortcomings of converting point clouds to meshes and spline surfaces are computational consuming and non-robustness. To overcome this, a direct tool path planning method (iso-parametric) for point clouds is presented. It follows a conformal point cloud parameterization laying a foundation for the planning. A simple mathematical formulation for determining steps analytically is then presented. Finally, tool path can be generated by iteratively computing the forward and side step and the linear interpolation. The angle preserving property of the conformal parameterization simplifies the calculation of tool path parameters as well as the transferring of surface interval to parametric step. What's more, the free-boundary property helps generate boundary conformed tool path.

However, this work preserves inherently the weakness of iso-parametric methods: the spatial path interval is uneven and thus less efficient than the iso-scallop method. Fortunately, many results of this paper hold for iso-scallop tool paths. The tool path planning for point clouds with arbitrary topology (e.g., a point cloud with holes) is not studied. Yet, by controlling the boundary map, tool path for them can be generated in a similar way to the presented.

## Acknowledgement

The authors are grateful for the support provided by National Key Basic Research Project of China (grant # 2011CB302400) and National Natural Science Foundation of China (grants # 50975274, 50975495).